\documentclass[fleqn,a4paper,12pt]{article}
\usepackage{amssymb}
\usepackage{amsmath}
\usepackage[dvips]{graphicx}
\usepackage{placeins}
\usepackage{lscape}
\usepackage{a4}
\usepackage{epsfig}

\newcommand{\p}{\partial}
\newcommand{\bea}{\begin{array}}
\newcommand{\eea}{\end{array}}

\newcommand{\be}[1]{\begin{equation}\label{#1}}
\newcommand{\bs}[1]{\begin{equation}\label{#1}\arraycolsep=0em\begin{array}{l}}
\newcommand{\es}{\end{array}\end{equation}}
\newcommand{\bss}{\[\arraycolsep=0em\begin{array}{l}}
\newcommand{\ess}{\end{array}\]}
\newcommand{\bc}{\begin{center} }
\newcommand{\ec}{\end{center} }
\newcommand{\bt}[1]{\begin{tabular}{#1}}
\newcommand{\et}{\end{tabular} }

\catcode`@=11 \long
\def\@caption#1[#2]#3{\par\addcontentsline{\csname
ext@#1\endcsname}{#1} {\protect\numberline{\csname
the#1\endcsname}{\ignorespaces #2}} \begingroup \small
\@parboxrestore \@makecaption{\csname fnum@#1\endcsname}
{\ignorespaces #3}\par \endgroup} \catcode`@=12

\catcode`@=11 \long
\def\@caption#1[#2]#3{\par\addcontentsline{\csname
ext@#1\endcsname}{#1} {\protect\numberline{\csname
the#1\endcsname}{\ignorespaces #2}} \begingroup \small
\@parboxrestore \@makecaption{\csname fnum@#1\endcsname}
{\ignorespaces #3}\par \endgroup} \catcode`@=12

\begin{document}

\allowdisplaybreaks
 \begin{titlepage} \vskip 2cm

\begin{center} {\Large\bf {Higher order symmetries} for linear and nonlinear  Schr\"odinger
equations}
\footnote{E-mail: {\tt nikitin@imath.kiev.ua} } \vskip 3cm {\bf {A.
G. Nikitin } \vskip 5pt {\sl Institute of Mathematics, National
Academy of Sciences of Ukraine,\\ 3 Tereshchenkivs'ka Street,
Kyiv-4, Ukraine, 01601\\}}\end{center} \vskip .5cm \rm

\begin{abstract}
We study arbitrary order symmetry operators for the linear
Schr\"odinger equations with arbitrary number of spatial variables.
We deduce determining equations for coefficient functions of such
operators and consider in detail some cases when these equations can
be explicitly solved. In addition, the complete group classification of the
nonlinear Schr\"odinger equation is presented.
\end{abstract}
\end{titlepage}
\section{Introduction}

Higher order symmetry operators present powerful tools for
investigations  of partial differential equations. They can be
used to construct constants of motion for evolution equations, to
search for coordinate systems in which solutions with separated
variables exist, and also for investigations of superintegrable
systems whose existence was predicted by Arnold Sommerfeld in 1923
\cite{S} while the systematical inquiry into the
superintegrability problem was started by P. Winternitz and
co-authors \cite{W}.

We remind that a nice kind of symmetries called {\it
supersymmetry} is also
connected with existence of higher order symmetry operators.

Investigations of symmetries of the Schr\"odinger equation (SE)
started with
papers of Niederer  and Boyer \cite{B} who presented group
classification of linear SEs with potential terms.
Such classification is connected with description of {\it first
order} symmetry operators for the related equation.

In fundamental paper \cite{P} the
complete description of {\it second order} symmetry operators for
the
1+2 dimensional SE was presented. Applications of these operators to the separation of variables
are discussed in the Miller book \cite{M}.

The problem of description of {\it arbitrary order} symmetry
operators for the Schrodinger equation was formulated in papers
\cite{J} and \cite{Fu}. In paper \cite{Ni} the complete set of
such operators was found for the case of trivial potential. In
\cite{FN} the third order symmetry operators for 1+1 dimensional
SE with arbitrary potential were described and used to construct
exact solutions. Moreover, a rather nontrivial connections between
the existence of the third order symmetry operators for the {\it
linear} SE and exact solutions of the related {\it nonlinear}
equation was established in \cite{FN}.

The very existence of many of exact solutions for the nonlinear SE
can be understood starting with its symmetries with respect to
continuous groups of transformations. The group analysis of the
nonlinear SE was carried out by a number of investigators, refer,
e.g., to \cite{Ch} and references cited therein. However, in fact
the related results were valid for ad hoc restricted classes of
the nonlinear SEs. The completed group classification of this
equation was presented in the recent paper \cite{Po}, the related
problem for systems of reaction-diffusion equations was solved in
\cite{R}.

In the present paper we deduce the determining equations for
coefficients of arbitrary order symmetry operators for the $1+m$
dimensional SE with arbitrary number $m$ of spatial variables. We
consider some cases when this system can be completely
solved, and discuss the related symmetry operators and their
possible applications. In addition, we present the complete group
classification of the nonlinear SE.

\section{Symmetry operators of arbitrary
order}
\subsection{Determining equations}
Consider the SE with an arbitrary number $m$ of spatial variables
\begin{equation}\label{2.1}
L\Psi(t,x)\equiv \left(i\frac{\partial}{\partial
t}-\frac{p^2}{2M}-V(x)\right)
\Psi(t,x)=0.
\end{equation}
Here -- $p^2=\Delta_m$ is the Laplace operator in space of $m$
variables, $X=(x_1, x_2, \ldots x_m)$, $ \Psi(t, {\bf x})$ is a
complex wave function.

For simplicity we restrict ourselves to such solutions of
(\ref{2.1}) which are defined on an open set $D$ of the
$m+1$-dimensional manifold $R$ consisting of points with
coordinates $(t, x_1, x_2, \ldots, x_m)$ and are analytic in
variables $t, x_1, \ldots, x_m$. The set of such solutions forms a
complex vector space $F_0$. Fixing $D$ (e.q., supposing that $D$
coincides with $R$) we come to space $F_0$ of solutions of the SE.

Let us denote by $F$ the vector space of complex valued functions
defined and real-analytic on $D$ and by $L$ the linear operator
(\ref{2.1}) defined on $F$. Then $L \in \Psi$ is $\Psi \in F$ and
$F_0$ is the subspace of $F$ which coincides with the zero-space
of $L$.

Let $Q$ be a n-order differential operator defined on $L$:
\begin{equation}\label{2.2}
{\tilde Q}_j=h^{a_1,a_2,\ldots, a_j}\frac{\partial^j} {\partial
x_{a_1} \partial x_{a_2}
\ldots \partial x_{a_j}}.
\end{equation}
Here and in the following the summation over the repeating indices is imposed over the values $1,2,..., m$.

We say $Q$ is a symmetry operators for SE (\ref{2.1}) provided
\begin{equation}\label{2.3}
[L,Q]=\alpha_Q L
\end{equation}
where $[\ ,\ ]$ denotes commutator and $\alpha_Q$ is a differential
operator of order ${n-1}$.

It is convenient to represent $Q$ as a sum of
$j$-multiple anticommutators \cite{FN}
 \begin{equation}\label{2.4}Q=\sum_{j=0}^{n}{Q}_j, \ \ Q_j=\left
[\left [\ldots \left[ K^{a_1a_2\ldots
a_j},\frac{\partial}{\partial x_{a_1}}\right]_+,
\frac{\partial}{\partial x_{a_2}}\right]_+,\ldots,
\frac{\partial}{\partial x_{a_j}}\right]_+\end{equation} where
$K^{a_1a_2\ldots a_j}$ are symmetric tensors depending on $t$ and
${\bf x}$ and $\left[K^{*},\frac{\partial}{\partial x_{a_j}}\right]=K^{*}\frac{\partial}{\partial x_{a_j}}+\frac{\partial}{\partial x_{a_j}}K^{*}$. Commuting (\ref{2.4}) with the Laplace operator  we automatically come to a sum of $j+1$ multiple anticommutators, and it is the reason why just this  representation of $Q$ leads to the most simple system of determining equations.

It is possible {\it ad hoc}  set in (\ref{2.3}) $\alpha_Q=0$
in as much as the term $\frac{\partial}{\partial t}$ cannot appear
in the l.h.s. of this equation.  Substituting (\ref{2.4}) into
(\ref{2.3}) and equating the coefficients for all powers of
differential operators we come to the following system of
determining equations
\begin{equation}\label{2.5}\begin{array}{l}
\partial^{(a_{n+1}}K^{a_1a_2\ldots a_n)}=0,\\
2{\dot K}^{a_1 a_2...a_{2s}}+ \frac1{M}
\partial^{(a_{2s}}K^{a_1 a_2...a_{2s-1})}\\
+ \sum\limits_{k=s}^{\left[(n-1)/2 \right]}(-1)^{s+k+1}{2(2k+1)!
\over
(2k-2s+1)!(2s)!}U_k^{a_1 a_2...a_{2s}}=0,\\
2{\dot K}^{a_1 a_2...a_{2l+1}}+\frac1{M}
\partial^{(a_{2l+1}}K^{a_1 a_2...a_{2l})}\\
+ \sum\limits_{k=l+1}^{\left[n/2 \right]}(-1)^{k+l}\frac{ 2(2k)!}
{(2k-2l-1)!(2l+1)!}W_k^{a_1 a_2...a_{2l+1}}=0 \end{array}
\end{equation}
where the dot denotes the derivative with respect to $t$, i.e., $\dot K^{...}=\frac{\partial K^{...}}{\partial t}$,
\begin{gather}\nonumber s=0,1,...,\left[n/2 \right],\quad
l=0,1,...,\left[(n-1)/2 \right],\\\begin{split}&
U_k^{a_1 a_2...a_{2s}}=K^{a_1 a_2...a_{2s}b_1
b_2...b_{2k-2s+1}}\partial_{b_1}\partial_{b_2}...\partial_{b_{2k-
2s+1}}
V,\\&
W_k^{a_1a_2...a_{2l+1}}=K^{a_1 a_2...a_{2l+1}b_1
b_2...b_{2k-2l-1}}\partial_{b_1}\partial_{b_2}...\partial_{b_{2k-
2l-1}}V\end{split}\label{nn1}\end{gather}
and the complete symmetrization is imposed over the indices in
brackets.

By definition, the terms with {\it negative} values of index numbers are identically zeros, while the index with number zero is forbidden. For example,
it is the case for $W_k^{a_1a_2...a_{2l+1}}$ if $k=l$, i.e., $W_k^{a_1a_2...a_{2k+1}}=0$, (see (\ref{nn1})),  and ${\dot K}^{a_1 a_2...a_{2s}}=\dot K$ if $s=0$.

The system  (\ref{2.5}) describes the coefficients of $n$-order symmetry operator for non-stationary Schr\"odinger equations. However, it is valid also for the stationary equations provided functions $K^{a_1a_2\ldots
a_j},\ \ j=1,2,...,n$ are time independent. It is interesting to note that in this case the system  (\ref{2.5}) is decoupled to two independent subsystems one of which includes tensors $K^{a_1a_2\ldots
a_j}$ with odd number of indices while the other is defined for even $j$. Thus the $n$-order integral of motion can be represented as a sum of two independent  operators (\ref{2.4}) of orders $n$ and $n-1$.  The first of them includes terms with $j=n,\ n-2,\ n-4,\ ...$ while for the second one we have expression (\ref{2.4}) with $j=n-1,\ n-3,\ n-5,\ ...$.

We see that integrals of motions with even and odd $d$ are linearly independent and can be considered separately. For even $n$ the determining equations (\ref{2.5}) are reduced to the following form:
\begin{gather}\label{new1}\begin{split}& \partial^{(a_{n+1}}K^{a_1a_2\ldots a_n)}=0,\quad n=2r,\\&
\partial^{(a_{2l+1}}K^{a_1 a_2...a_{2l})}
+ M\sum\limits_{k=l+1}^{r}(-1)^{k+l}\frac{ 2(2k)!}
{(2k-2l-1)!(2l+1)!}W_k^{a_1 a_2...a_{2l+1}}=0,\\&l=0,1,2,...,r-1\end{split}\end{gather}
while for $n$ odd we have:
\begin{gather}\label{new3}\begin{split}& \partial^{(a_{n+1}}K^{a_1a_2\ldots a_{n})}=0,\quad n=2r+1,\\&
\partial^{(a_{2s}}K^{a_1 a_2...a_{2s-1})}
+ M\sum\limits_{k=s}^{r}(-1)^{s+k+1}{2(2k+1)!
\over
(2k-2s+1)!(2s)!}U_k^{a_1 a_2...a_{2s}}=0,\\&s=0, 1,2,...,r.\end{split}\end{gather}

Let us stress that formulae (\ref{new1}) and (\ref{new3}) together with definitions (\ref{nn1}) present explicitly the determining equations for potentials and the related coefficients of integrals of motion of {\it arbitrary order $n$} and for {\it arbitrary number $m$} of spatial variables. We believe that the direct use of these equations could be convenient in studies of higher (and even arbitrary) order integrals of motion for quantum mechanical systems, which are rather popular now, see survey \cite{wi1} and paper \cite{wi2}. Indeed, the derivation of the determining equations for higher order symmetry operators with fixed $n$ and $m$  looks as unnecessary waste of time and energy since this job has been already done.

\subsection{Complete set of symmetry operators of arbitrary order}
For the case of {\it ad hoc} fixed potential $V(t,{\bf x})$ we can
try to find exact solutions of (\ref{2.5}) for {\it arbitrary} $n$.

If $V({\bf x})=0$ than the system of equations (\ref{2.5}) is
significantly  simplified and reduced to the following form

\begin{equation}\label{2.6}\begin{array}{l}
\partial^{(a_{j+1}}K^{a_1 a_2 ...a_j)}= -
2M{\dot K}^{a_1 a_2...a_{j+1}},\quad j=0,1,...n-1,\\
\partial^{(a_{n+1}} K^{a_1 a_2 ...a_n)} =0,\\{\dot K}=0,\quad
j=0.\end{array} \end{equation}

Equations (\ref{2.6}) can be solved explicitly for arbitrary $n$.
A differential consequence of (\ref{2.6}) is

\begin{equation}\begin{array}{l}\partial^{(a_{j+1}}\partial^{a_{j
+2}}...\partial
^{a_{j+s}}K^{a_1 a_2 ...a_j)}=0,\quad s=n-j+1.\end{array}
\label{2.7}\end{equation}

Solutions of (\ref{2.7}) are the generalized Killing tensors of
order $s$ and rank $j$ whose explicit form is presented in
\cite{nik}. The first of relations (\ref{2.6}) reduces to the
first order ordinary equations for arbitrary parameters defining
the generalized Killing tensors which are easily integrated. The
number $N_n$ of linearly independent $n$-order symmetries is given
by the following formula

$$\begin{array}{l}N_n=\frac1{ 4!}(n+1)(n+2)^3(n+3)
\end{array}$$
and all these symmetries belong to the enveloping algebra of the
Lie algebra of the Schr\"{o}dinger group \cite{Ni}.

\subsection{Third order symmetry operators for $1+1$-dimensional\\
Schr\"odinger equation}

One more simplified version of the determining equations
(\ref{2.5}) corresponds to a non-trivial potential in the
important case of the only spatial variable. Then the first and
second order symmetry operators reduce on the set of solutions of
the related SE to generators of a Lie group.

     Let us consider the case $n=3$, which
corresponds to the simplest non-Lie symmetry, in more detail. For
simplicity we set $m=1$ and $V=\frac{U}{2}$. The related tensors
$K^{a_1a_2...a_j}$ in (\ref{2.4}) reduce to scalars and so
\begin{equation}\begin{array}{l}\label{00}
Q=\left[\left[\left[h_3,\frac{\partial}{\partial x}\right]_+,
\frac{\partial}{\partial x}\right]_+,\frac{\partial}{\partial
x}\right]_+
\left[\left[h_2,\frac{\partial}{\partial
x}\right]_+,\frac{\partial}{\partial
x}\right]_++\left[h_1,\frac{\partial}{\partial
x}\right]_++h_0.\end{array}\end{equation}

The corresponding system (\ref{2.5}) reduces to
\begin{equation}\begin{array}{l}  h_3'=0,\quad \quad  h_2'+2
\dot{h}_3=0, \\2 \dot {h}_2 +h_1'
-
6h_3 U' =0,\end{array}\label{2.8} \end{equation}
\begin{equation}\label{2.9} \begin{array}{l} 2 \dot{h}_1+ h_0'-4h_2
U'=0,\\ \dot{h}_0-h_1U'+h_3U'''=0,
\end{array} \end{equation}
where the dots and primes denote derivatives w.r.t. $t$
and $x$ respectively.

Excluding $h_0$ from (\ref{2.9}) and using (\ref{2.8}) we arrive at
the
following equation
\begin{equation} \begin{array}{l}
aU''''-(2 \ddot{a} x^2+6aU+c-2 \dot{b}
x)U''-
\\ -6(2 \ddot{a} x +a U'- \dot{b})U'-12 \ddot{a} U-2(2
\partial _t^4 ax^2-2 \stackrel{\dots}{b} x+
\ddot{c})=0\end{array} \label{2.10}\end{equation}
where $a, b, c$ are arbitrary functions of $t$.
     Equation (\ref{2.10}) is nothing but the compatibility
condition for system (\ref{2.8}), (\ref{2.9}). If the potential $U$
satisfies
(\ref{2.10}) then the corresponding coefficients of the $SO$ have
the
form
\begin{equation} \begin{array}{l} h_3=a,\quad h_2=-2 \dot{a}
x+b,\quad h_1=g_1+6aU,\\h_0=-{4
\over
3} \stackrel{\dots}{a} x^3+2 \ddot{b} x^2-2 \dot{c} x-4 \dot{a}
\varphi+4(b-2 \dot{a} x)U+d \end{array} \label{2.11}\end{equation}
where
$$ \begin{array}{l} g_1=2 \ddot{a}
x^2-2 \dot{b} x+c,\quad \varphi= \int
Udx,\quad u=\varphi ',\qquad d=d(t). \end{array}   $$

Separating variables in (\ref{2.10}) we conclude that
up to
equivalence, this equation can be reduced to one of the following
forms:
\begin{equation} \begin{array}{l}
U''-3U^2+3\omega_1=0,\end{array} \label{2.13} \end{equation}
\begin{equation} \begin{array}{l}
U''-3U^2-8\omega_2x=0,\end{array} \label{2.14} \end{equation}
\begin{equation} \begin{array}{l}
(U''-3U^2)'-2\omega_3(xU'+2U)=0,\end{array} \label{2.15}
\end{equation}
\begin{equation} \begin{array}{l}
\varphi'''-3(\varphi')^2-2\omega_4(x^2 \varphi)'={1 \over 3}\omega_4^2
x^4+\omega_5\end{array}, U=\varphi'\label{2.16}\end{equation}
where $\omega_1,\dots \omega_5$ are arbitrary constants.

     From (\ref{00}), (\ref{2.13})-(\ref{2.16}) we find the
corresponding symmetry operators
\begin{equation}\label{2.161}\begin{array}{l} Q=p^3+{3 \over 4}\{
U,p
\}
\equiv 2pH+{1 \over
2} U p+ {i \over
4}U', \end{array} \end{equation}
\begin{equation} \begin{array}{l} Q=p^3+{3 \over 4}\{ U,p \}
-\omega_2t, \end{array} \label{2.17}\end{equation}
\begin{equation} \begin{array}{l} Q=p^3+{3 \over
4} \{ U,p \}+\omega_3 \left( tH -{1 \over 4} \{x,p \}
\right),\end{array} \label{2.18}\end{equation}
\begin{equation} \begin{array}{l} Q_{\pm}=\frac{1}{ \sqrt {24}}
\left[p^3 \pm {i \over
4}\omega \{ \{x,p \} ,p \} +{1\over 4} \{ 3 \varphi' -\omega^2x^2,p
\} \pm \right.\\
\left.\pm {i \over 2}\omega \left(\varphi+2x
\varphi'-{\omega^2 \over 3} x^3 \right) \right]\exp(\pm i\omega
t),\quad
\omega= \sqrt {-\omega_4}
\end{array}\label{2.19}\end{equation}
where $U$  are solutions of (\ref{2.13})-(\ref{2.16}) and
$H=\frac12 \left( -\frac{\partial}{\partial_x^2} +U(x)\right)$.

     Thus, {\it the Schr\"odinger equation (2.1) admits a
third-order $SO$ if potential $U$ satisfies one of the equations
(\ref{2.13})- (\ref{2.16})}. The explicit form of the
corresponding $SOs$ is present in (\ref{2.161})-(\ref{2.19}).

Formula (\ref{2.13}) presents the Weierstrass equation while
relation (\ref{2.14}) defines the first Painlev\'e transcendent.
Using  generalized Miura ansatz  equations (\ref{2.15}) and
(\ref{2.16}) can be reduced to the second and forth Painlev\'e
transcendents respectively or to special Riccatti equations
\cite{FN}.

Operators  (\ref{2.161})-(\ref{2.19}) together with Hamiltonian $
H$ form rather interesting algebras whose analysis lies out of
frames of the present paper.
\section{Group classification of the nonlinear SE}

It was shown in \cite{FN} that if the linear one dimensional SE
(\ref{2.1}) admits the third order symmetry operator (\ref{2.16})
than there exists a wide class of solutions for (\ref{2.16}) which
solve also the nonlinear SE with cubic nonlinearity. Thus there
exist rather non-trivial connections between the third order
symmetries for the linear SE and symmetries of the nonlinear SE
which cause the existence of exact solutions.

Here we present a complete description of classical Lie symmetries
of the non-linear Schr\"odinger equation
\begin{equation}\label{3.1}
\left(\frac{\partial}{\partial
t}+\Delta\right)\Psi+F(\Psi,\Psi^*)=0\end{equation} where
$\Psi=\Psi(t,{\bf x}), {\bf x}\in R_m$, $F$ is an arbitrary
function of two variables: $\Psi$ and complex conjugated function
$\Psi^*$.

Classical Lie symmetries of some special classes of equations
(\ref{3.1}) were investigated in numerous papers, refer, e.g., to
\cite{Ch} and references cited therein. Here we present
the results of complete group classification for all nonequivalent
nonlinearities $F$, based on results of papers \cite{Po} and
\cite{R}.

To describe symmetries of the SE with non-fixed nonlinearity in the l.h.s.
we can use either the classical Lie algorithm or its specific simplified version
\cite{R} which presupposes solution of the following operator equation
(compare with (\ref{2.1}))
$$ \left [L, Q\right] =\alpha_QL +\varphi(t, {\bf x})$$
where $Q$ is a first order differential operator, $\varphi$ is unknown function.
We will not present here details of calculations but present the final result of
group classification of (\ref{3.1}).

For arbitrary $F$ equation  (\ref{3.1}) is invariant with respect
the Euclid group $E(1,m)$ and so admits $1+\frac{m(m+1)}{2}$
symmetry operators of first order
\begin{equation}\begin{array}{l}\label{3.2}
P_0=\frac{\partial}{\partial t}, \ P_a=\frac{\partial}{\partial
x_a},  \\
J_{ab}=x_ap_b-x_bp_a, \ \ a,b=1,2,\dots ,
m.\end{array}\end{equation}

Al cases when the basic invariance group $E(1,m)$ can be extended
are enumerated in the following tables.

\begin{center}
{\normalsize {\bf Table 1.} Nonlinearities dependent on arbitrary function $f(\Omega)$ .
\par}\end{center}

\bc\renewcommand{\arraystretch}{1.6}\tabcolsep=2mm
\begin{tabular}{|p{7mm}|p{62mm}|p{21mm}|p{53mm}|}
\hline
&\hfil$F$\hfil&\hfil$\Omega$\hfil&\hfil Additional symmetries\hfil \\ \hline
$\hfil1.1$ & $f(\Omega)|\psi|^{\gamma_1}e^{\gamma_2\varphi}\psi$,
$\:\gamma_1^2+\gamma_2^2\not=0$ &
$|\psi|^{\gamma_2}e^{-\gamma_1\varphi}$ & $(\gamma_1^2+\gamma_2^2)D-\gamma_1I-\gamma_2M$
\\ \hline
$\hfil1.2$ & $(f(\Omega)+(\gamma-i)\delta\ln|\psi|)\psi$ & $|\psi|^\gamma e^{-\varphi}$ & $e^{\delta t}(I+\gamma M)$
\\ \hline
$\hfil1.3$ & $(f(\Omega)+\delta\varphi)\psi$, $\:\delta\not=0$&
$|\psi|$ & $e^{\delta t}M$, $\;e^{\delta t}(\p_a+\frac{1}{2}\delta x_aM)$
\\ \hline
$\hfil1.4$& $f(\Omega)\psi$ & $|\psi|$ & $M$, $\;G_a$
\\ \hline
$\hfil1.5$& $f(\Omega)e^{i\psi}$ & $\Re\psi$ & $D+i(\p_\psi-\p_{\psi^*})$
\\ \hline
$\hfil1.6$& $f(\Omega)+i(\delta_1+i\delta_2)\psi$ & $\Re\psi$ &
$ie^{-\delta_1t}\theta(x)(\p_\psi-\p_{\psi^*})$
\\ \hline
\end{tabular}\ec

\vspace{1ex}
Here $\gamma$, $\gamma_1$, $\gamma_2$, $\delta$, $\delta_1$, $\delta_2$~ are real numbers,
$\theta=\theta(x)\ \in\ R$~ is a solution of equation $\:\Delta\theta=\delta_2\theta$,
We present our results in terms of  amplitude $\rho=|\psi|$
and phase~$\varphi=\frac{i}{2}\ln\frac{\psi^*}{\psi}$ of function
~$\Psi$ and use the following notations
\begin{gather*}\begin{split}&
I:=\Psi\p_\Psi+\Psi^*\p_{\Psi^*}=\rho\p_\rho, \quad
M:=i(\Psi\p_\Psi-\Psi^*\p_{\Psi^*})=\p_{\varphi}, \\&
D:=t\p_t+\dfrac{1}{2}x_a\p_a, \quad G_a:=t\p_a+\dfrac{1}{2}x_aM, \\&
\Pi:=t^2\p_t+tx_a\p_a-\dfrac{n}{2}tI+\dfrac{1}{4}x_ax_aM.
\end{split}\end{gather*}


In the following page we present non-linearities which are defined up to arbitrary parameters.


\begin{center}
{\normalsize {\bf Table 2.} Nonlinearities depending on arbitrary parameters.
\par}\end{center}

\bc\renewcommand{\arraystretch}{1.6}\tabcolsep=2mm
\normalsize\begin{tabular}{|p{8mm}|p{54mm}|p{83mm}|}
\hline
&\hfil$F$\hfil&\hfil Additional symmetries\hfil
\\ \hline
$\hfill 2.1$ & 0 & $G_a$, $\:I$, $\:M$, $\:D$, $\:\Pi$, $\:\eta^0\p_\psi+{\eta^0}^*\p_{\psi^*}$
\\ \hline
$\hfill 2.2$ & $\gamma\psi+\psi^*$ & $I$, $\:\eta^0\p_\psi+{\eta^0}^*\p_{\psi^*}$
\\ \hline
$\hfill 2.3$ & $\sigma|\Re\psi|^\gamma,$ $\gamma\not=0,1$
 & $I+(1-\gamma)D$, $\; i\theta(x)(\p_\psi-\p_{\psi^*})$
\\ \hline
$\hfill 2.4$ & $\sigma\ln|\Re\psi|$&
$I+D-i(t\Re\sigma+\frac{1}{2n}x_ax_a\Im\sigma)(\p_\psi-\p_{\psi^*})$, \hfill ${}$
$\;i\theta(x)(\p_\psi-\p_{\psi^*})$
\\ \hline
$\hfill 2.5$ & $\sigma e^{\Re\psi}$ &
$D-\p_\psi-\p_{\psi^*}$, $\;i\theta(x)(\p_\psi-\p_{\psi^*})$
\\ \hline
$\hfill 2.6$ & $\sigma|\psi|^{\gamma_1}e^{\gamma_2\varphi}\psi$, \quad $\gamma_2\not=0$ &
$M-\gamma_2D$, $\;\gamma_2I-\gamma_1M$
\\ \hline
$\hfill 2.7$ & $\sigma|\psi|^{\gamma}\psi$,
\quad $\gamma\not=0,\frac{4}{n}$& $G_a$, $\;M$, $\;I-\gamma D$
\\ \hline
$\hfill 2.8$ & $\sigma|\psi|^{4/n}\psi$, & $G_a$, $\;M$, $\;I-\frac{4}{n}D$, $\;\Pi$
\\ \hline
\multicolumn{3}{|c|}{In he following \quad $F=(-(\delta_1+i\delta_2)\ln|\psi|+(\delta_3-i\delta_4)\varphi)\psi$,\quad
$\Delta=(\delta_2-\delta_3)^2-4\delta_1\delta_4$}
\\ \hline
$\hfill 2.9$ & $\delta_4=0$, $\delta_3\not=0$, $\delta_2\not=\delta_3$ &
$e^{\delta_3t}M$, $\;e^{\delta_3t}(\p_a+\frac{1}{2}\delta_3x_aM)$,
$\;e^{\delta_2t}(I-\frac{\delta_1}{\delta_2-\delta_3}M)$
\\ \hline
$\hfill 2.10$ & $\delta_4=0$, $\delta_3\not=0$, $\delta_2=\delta_3$ &
$e^{\delta_3t}M$, $\;e^{\delta_3t}(\p_a+\frac{1}{2}\delta_3x_aM)$,
$\;e^{\delta_2t}(I-\delta_1tM)$
\\ \hline
$\hfill 2.11$ & $\delta_4=0$, $\delta_3=0$, $\delta_2\not=0$ &
$M$, $\;G_a$, $\;e^{\delta_2t}(\delta_2I-\delta_1M)$
\\ \hline
$\hfill 2.12$ & $\delta_4=0$, $\delta_3=0$, $\delta_2=0$, $\delta_1\not=0$ &
$M$, $\;G_a$, $\;I-\delta_1tM$
\\ \hline
$\hfill 2.13$ & $\delta_4\not=0$, $\Delta>0$ &
$e^{\lambda_it}(\delta_4I+(\lambda_i-\delta_2)M)$, $i=1,2$, \hfill ${}$
$\lambda_1=\frac{1}{2}(\delta_2+\delta_3-\sqrt{\Delta})$,
$\lambda_2=\frac{1}{2}(\delta_2+\delta_3+\sqrt{\Delta})$
\\ \hline
$\hfill 2.14$ & $\delta_4\not=0$, $\Delta<0$ &
$e^{\mu t}(\delta_4\cos\nu t\,I+((\mu-\delta_2)\cos\nu t-\nu\sin\nu t)M)$, \hfill ${}$
$e^{\mu t}(\delta_4\sin\nu t\,I+((\mu-\delta_2)\sin\nu t+\nu\cos\nu t)M)$, \hfill ${}$
$\mu=\frac{1}{2}(\delta_2+\delta_3)$, $\:\nu=\frac{1}{2}\sqrt{-\Delta}$
\\ \hline
$\hfill 2.15$ & $\delta_4\not=0$, $\Delta=0$ &
$e^{\mu t}(\delta_4tI+\frac{1}{2}(\delta_3-\delta_2)tM+M)$, \hfill ${}$
$e^{\mu t}(\delta_4I+\frac{1}{2}(\delta_3-\delta_2)M)$, 
$\;\mu=\frac{1}{2}(\delta_2+\delta_3)$
\\ \hline
\end{tabular}\ec

Here  $\gamma$, $\gamma_1$, $\gamma_2$,
$\delta_1$, $\delta_2$, $\delta_3$, $\delta_4$~ are real numbers,
 $\sigma\not=0$ ;
$\eta^0$~ is an arbitrary solution of the initial equation,
$\theta=\theta(x)\in R$~ is a solution of the Laplace equation $\Delta\theta=0.$

Thus we present the results of group classification of the nonlinear Shr\"odinger
equation. We will not discuss possible applications of the obtained results which
consist in description all possible models with non-trivial extention of the basic invariance
with respect to the group $E(1,n)$.

We stress that our classification is valid for nonlinear SE with {\it arbitrary} number of spatial
variables. Rather surprisingly, increasing of the number of these variables do not lead to increasing of calculations difficulties.

\section{Note added to this preprint version}

This preprint includes the contribution to the CRM Proceedings \cite{NNN} which is slightly corrected and added by the text placed in the first half of Page 3. The reasons of its publication is caused by the current interest in higher and arbitrary order integrals of motion, see, e.g., refs.  \cite{wi1} and \cite{wi2}. I believe that the determining equations for such integrals, valid for arbitrary number of independent variables, could be interesting for researcher working in this field. Just these equations were presented in \cite{NNN} for more general case of non-stationary quantum mechanical systems. And the additional text on Page 3 presents these equations reduced to the case of stationary systems.

\bibliographystyle{amsalpha}

\end{document}